\documentclass[aps,prl,superscriptaddress,twocolumn]{revtex4} 
\usepackage{graphicx}
\usepackage{epstopdf}

\begin{document}

\newcommand{\etal}{{\it et al.}\/}
\newcommand{\gtwid}{\mathrel{\raise.3ex\hbox{$>$\kern-.75em\lower1ex\hbox{$\sim$}}}}
\newcommand{\ltwid}{\mathrel{\raise.3ex\hbox{$<$\kern-.75em\lower1ex\hbox{$\sim$}}}}

\title{Pairfield Fluctuations of the 2D Hubbard Model}

\author{T.~A.~Maier}
\affiliation{Computational Sciences and Engineering Division, Oak Ridge
National Laboratory, Oak Ridge, Tennessee 37831-6164, USA}

\author{D.~J.~Scalapino}
\affiliation{Department of Physics, University of California, Santa Barbara,
CA 93106-9530, USA}

\date{\today}

\begin{abstract}
At temperatures above the superconducting transition temperature the pairfield
susceptibility provides information on the nature of the pairfield
fluctuations. Here we study the $d$-wave pairfield susceptibility of a 2D
Hubbard model for dopings which have a pseudogap (PG) and for dopings which do
not. One knows that in both cases there will be a region of
Kosterlitz-Thouless fluctuations as the transition at $T_{\rm KT}$ is
approached. Above this region we find evidence for Emery-Kivelson phase
fluctuations for dopings with a PG and Gaussian amplitude fluctuations for
dopings without a PG.
\end{abstract}


\maketitle



Tunneling experiments have been used to study pairfield fluctuations in both
the underdoped and overdoped cuprates \cite{ref:5,ref:Koren}. In these
experiments the tunneling current $I$ versus voltage $V$ between an optimally
doped YBCO ($T_c^{\rm High}\sim90$ K) electrode and an underdoped or overdoped
($T_C^{\rm Low}\sim50$ K) electrode was measured. The change in the $I$-$V$
characteristic $\Delta I(V)$ with the application of a small magnetic field or
under microwave irradiation, which suppress the pairfield current, gives the
contribution associated with the transfer of pairs from the higher $T_c$
electrode to the fluctuating pairfield of the lower $T_c$ electrode. Similar
phenomena are well-known in the traditional low $T_c$ superconductors
\cite{ref:4} where the fluctuating pairfield is well described by
time-dependent Ginzberg-Landau (TDGL) theory \cite{ref:2} with parameters set
by the  lattice phonon spectrum and the Fermi liquid out of which the
superconducting state emerges. In the case of the high $T_c$ cuprates,
$T_c/E_F$ is larger, the materials are quasi two-dimensional and  depending
upon the doping, the superconducting phase can emerge from a pseudogap (PG)
phase or a non-pseudogap phase.


Various authors \cite{ref:Janko,ref:Kwon,ref:She} have discussed the
possibility of using pair tunneling as a probe to study the differences in the
pairfield fluctuations between the PG and non-PG regions. Here, after defining
the pairfield susceptibility and describing the type of experiment which
motivated this study, we use the dynamic cluster approximation (DCA) with a
continuous time auxiliary-field quantum Monte Carlo (QMC) solver to study the
pairfield fluctuations for a 2D Hubbard model with an onsite Coulomb
interaction $U/t=7$ and a next-nearest-neighbor hopping $t'/t=-0.15$ in units
of the nearest-neighbor hopping amplitude $t$. Previous calculations have
shown that in the underdoped regime this model exhibits a peak in the spin
susceptibility \cite{ref:M,ref:C} and an antinodal gap in the single particle
spectral weight \cite{ref:a,ref:b} characteristic of a PG. Our aim is to
compare the nature of the pairfield fluctuations as the superconducting phase
is approached for fillings which exhibit a PG with fillings which do not.

The dynamic $d$-wave pairfield susceptibility $\chi_d(\omega,T)$ is given by the
Fourier transform of the pairfield response function
\begin{equation}
  \chi_d(t,T)=-i\left\langle\left[\Delta_d(t),\Delta^\dagger_d
  (0)\right]\right\rangle\theta(t)
\label{eq:1}
\end{equation}
with
\begin{equation}
  \Delta^\dagger_d=\frac{1}{\sqrt N}\sum_k(\cos k_x-\cos
  k_y)c^\dagger_{k\uparrow}c^\dagger_{k\downarrow}
\label{eq:2}
\end{equation}
In the ladder or TDGL approximation \cite{ref:2}
\begin{equation}
  \chi_d(\omega,T)\sim\frac{1}{\varepsilon(T)-i\frac{\omega}{\Gamma_0}}
\label{eq:3}
\end{equation}
with
\begin{equation}
  \varepsilon(T)=\ln\left(\frac{T}{T_c}\right)\simeq\frac{T-T_c}{T_c}
\label{eq:4}
\end{equation}
and $\Gamma_0=8T_c/\pi$.
\begin{figure}[htbp]
\includegraphics[width=7.5cm]{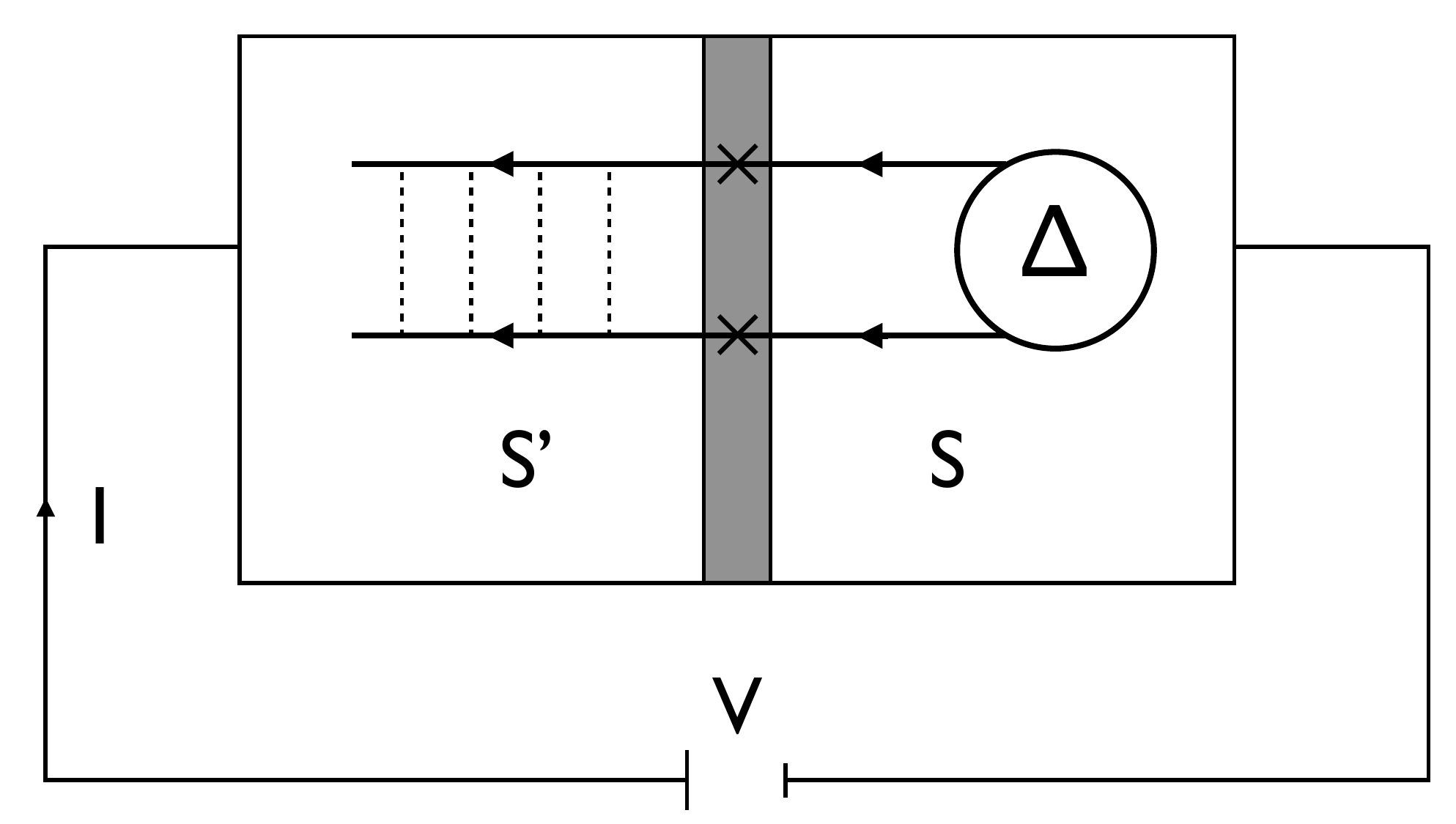}
  \caption{Illustration of a pair tunneling experiment showing a tunnel junction
	separating two  films $S$ and $S'$. The temperature $T$ is such that it is
	lower than the transition temperature $T_c$ of the film $S$ on the right
	and higher than the transition temperature $T'_c$ of the film $S'$ on the
	left. In this case there will be an excess current associated with
	electron pairs from $S$ tunneling to the fluctuating pairfield of
	$S'$.\label{fig:1}}
\end{figure}
As schematically illustrated in Fig.~\ref{fig:1}, tunneling experiments
\cite{ref:5,ref:Koren,ref:4} between a superconducting film $S$ below its
transition temperature and a film $S'$ above its transition temperature find
an excess current. This excess current is associated with the transfer of
pairs from the superconducting side to the fluctuating pairfield on the
non-superconducting side \cite{ref:6,ref:7}. This current varies as ${\rm
Im}\,\chi(\omega=2eV)$, which for the TDGL form Eq.~(\ref{eq:3}) gives
\begin{equation}
  \Delta I(V)\sim\frac{\left(\frac{2eV}{\Gamma_0}\right)}{\varepsilon^2(T)+\left
  (\frac{2eV}{\Gamma_0}\right)^2}\,.
\label{eq:5}
\end{equation}
The temperature dependence of the peak in $\Delta I(V)$ at
$2eV=\Gamma_0\varepsilon(T)$ provides information on the nature of the
pairfield fluctuations on the non-superconducting side. 
Independent of the TDGL approximation, in general the voltage integral of
$\Delta I(V)/V$ is proportional to $\chi_d(\omega=0,T)$. 

\begin{figure}[t!]
\includegraphics[width=8.5cm]{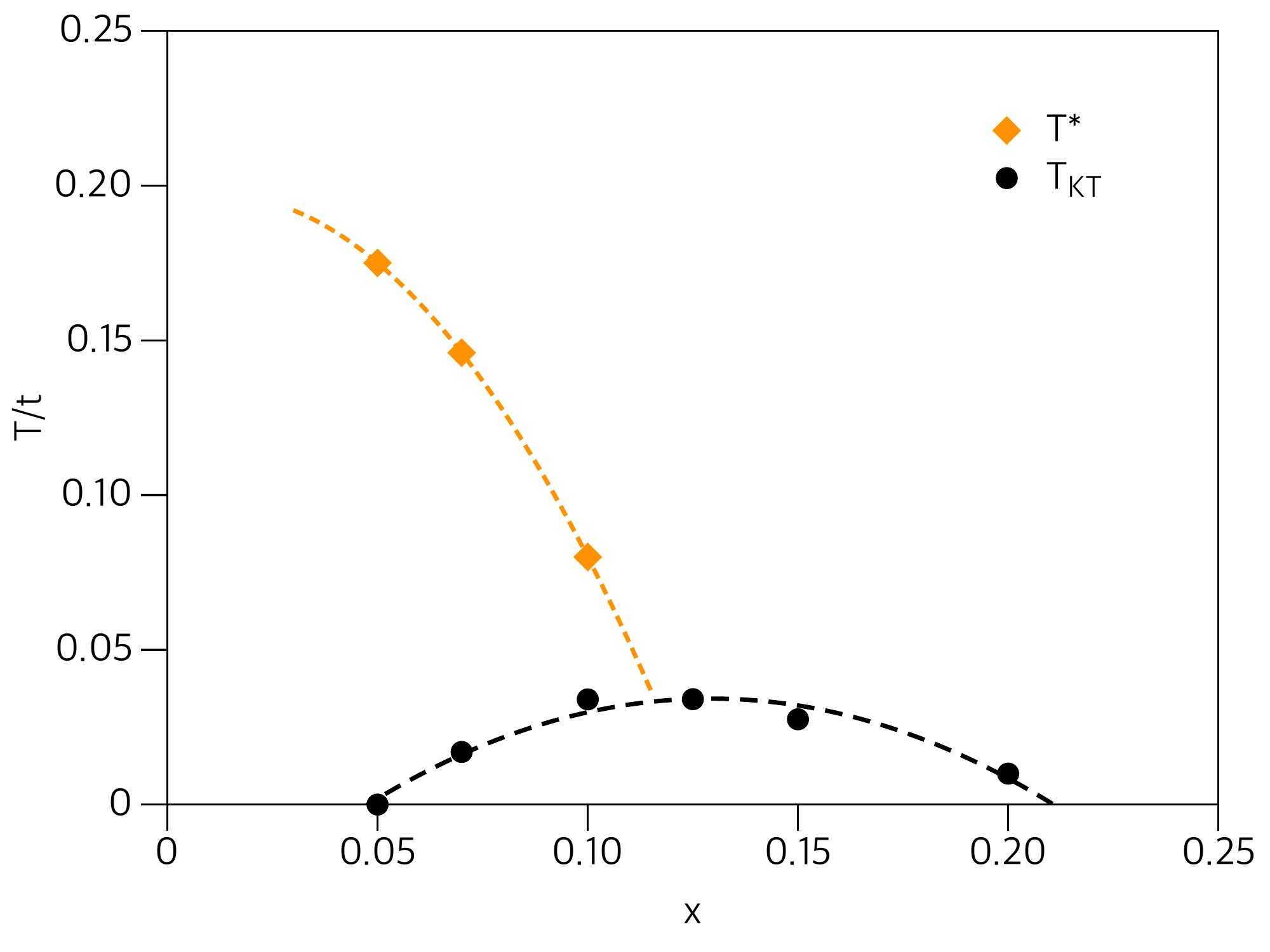}
  \caption{(Color online) Schematic temperature-doping phase diagram of the 2D
  Hubbard model.
  There is long-range AF order at $T=0$ for $n=1$, a superconducting region
  with a Kosterlitz-Thouless \protect\cite{ref:3} transition line labeled by
  $T_{\rm KT}$ and a dashed pseudogap (PG) line labeled $T^*$ where the $q=0$
  spin susceptibility peaks. The orange diamonds mark the peaks of the spin
  susceptibility shown in Fig.~\protect\ref{fig:3}, and the black circles mark
  the temperature at which the extrapolation of the $d$-wave eigenvalue
  $\lambda_d(T)$ reaches one. \label{fig:2}}
\end{figure}

Such experiments require a careful choice of materials and special fabrication
techniques. In addition, the measurements are limited to temperatures below
the $T_c$ of the higher transition temperature film and require the careful
separation of the excess pair current from the quasi-particle background.
Here, motivated by these experiments, we will carry out a numerical study of
the $d$-wave pairfield fluctuations. While this will not have the same
limitations as the experiment, it is limited by our choice of the 2D Hubbard
model, by the DCA approximation and by the fact that the simulation works with
Matsubara frequencies.  As we will see, this basic model exhibits features
seen in the cuprate materials and the DCA approximation allows us to go beyond
the TDGL result. In addition, we will avoid the problem of analytic
continuation of the Matsubara frequencies by calculating the $\omega=0$ response,
which as noted is proportional to the voltage integral of $\Delta I(V)/V$.

The $d$-wave pairfield susceptibility that we will study is given by
\begin{equation}
\chi_d(T)=\frac{\chi_{d0}(T)}{1-\lambda_d(T)}
\end{equation}
with
\begin{equation}
\chi_{d0}(T)=\frac{T}{N}\sum_k\phi^2_d(k)G(k)G(-k)
\end{equation}
Here $G(k)$ is the dressed single particle propagator and $\lambda_d(T)$ and
$\phi_d(k)$ are the $d$-wave eigenvalue and eigenfunction obtained from the
Bethe-Salpeter equation
\begin{equation}
  -\frac{T}{N}\sum_{k'}\Gamma_{pp}(k,k')G(k')G(-k')\phi_d(k')=\lambda_d\phi_d(k)
\label{eq:7}
\end{equation}
with $\Gamma_{pp}$ the irreducible particle-particle vertex. The notation
$k$ denotes both the momentum {\bf k} and Fermion Matsubara frequency
$\omega_n=(2n+1)\pi T$. These quantities are evaluated using a DCA QMC
approximation \cite{ref:8}, in which the momentum space is coarse-grained to
map the problem onto a finite size cluster embedded in a mean-field, which
represents the lattice degrees of freedom not included on the cluster. The
effective cluster problem is then solved with a continuous-time
auxiliary-field QMC algorithm \cite{ref:GullEPL}. Here, we use a 12-site
cluster (see Fig.~1 in Ref.~\cite{ref:Maier05PRL}), which allows us to study
the effects of non-local fluctuations, and for which the Fermion sign poblem
of the QMC solver is manageable down to temperatures close to the
superconducting instability.

\begin{figure}[t!]
\includegraphics[width=8.5cm]{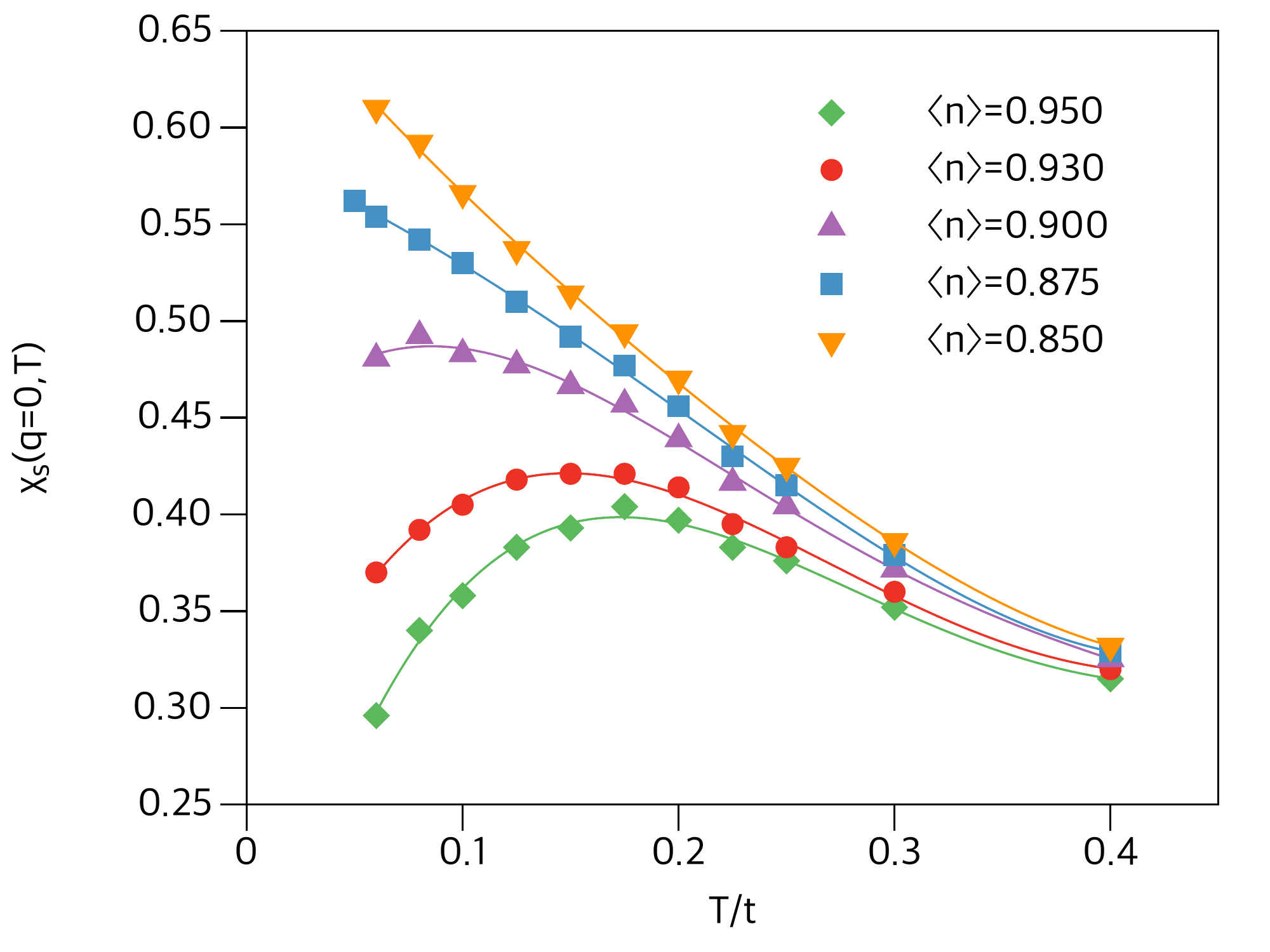}
  \caption{(Color online) The $q=0$ spin susceptibility $\chi_s(T)$ versus
  temperature for the
  various dopings. At the smaller dopings, $\chi_s(T)$ exhibits a peak indicating
  the opening of a PG. 
  \label{fig:3}}
\end{figure}

A schematic temperature-doping phase diagram estimated from these calculations
is shown in Fig.~\ref{fig:2}.
\begin{figure*}[t!]
\includegraphics[width=15cm]{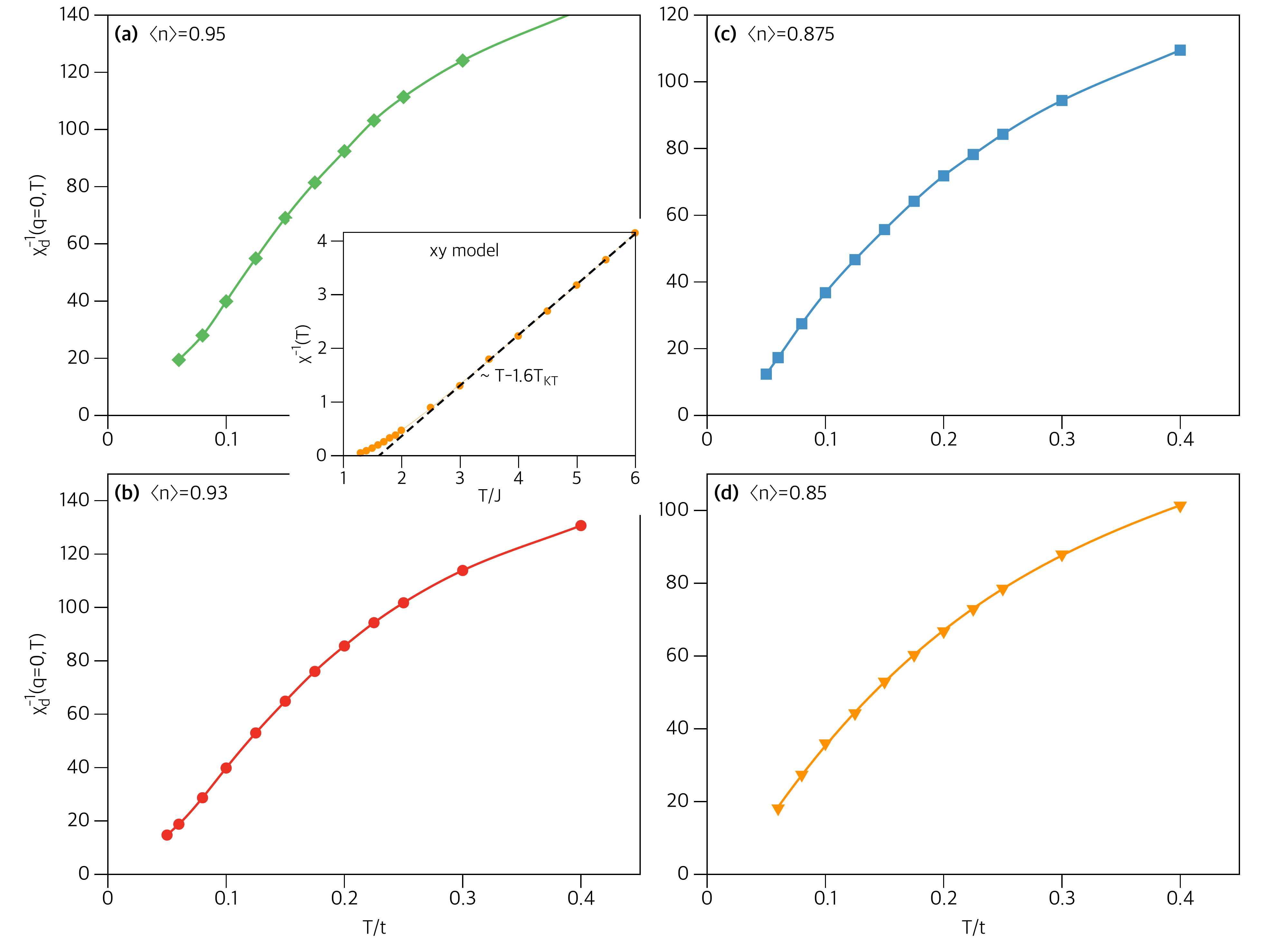}
  \caption{Plots of $\chi^{-1}_d(T)$ for the 2D Hubbard model versus
  temperature for various fillings. The inset shows Monte Carlo results for
  the susceptibility of a 2D xy model, which has a fixed amplitude with only a
  phase degree of freedom that can fluctuate.
  \label{fig:4}}
\end{figure*}
At half-filling the groundstate has long-range AF order which is absent at
finite temperature because of the continuous rotational spin symmetry. For low
doping, DCA \cite{ref:C,ref:M} calculations find a peak in the $q=0$
spin susceptibility $\chi_s(T)$. There is also evidence for the opening of an
antinodal gap in the single particle spectral weight \cite{ref:a,ref:b} as the
temperature drops below $T^*$. Results for $\chi_s(T)$ for $U/t=7$ and
$t'/t=-0.15$ are shown in Fig.~\ref{fig:3}. For $\langle n\rangle=0.95$ and
$\langle n\rangle=0.93$ the spin susceptibility $\chi_s(T)$ peaks and then
decreases below a temperature $T^*$ which marks the opening of a pseudogap.
The behavior of $\chi_s(T)$ for $\langle n\rangle=0.875$ and 0.85 are
consistent with dopings that are beyond the PG region.

At lower temperatures dynamic cluster calculations also find evidence for
$d$-wave superconductivity \cite{ref:Maier05PRL,ref:Staar14}, which for a 2D
system will occur at a Kosterlitz-Thouless \cite{ref:3} transition $T_{\rm
KT}$. Here we are interested in comparing the manner in which the pairfield
fluctuations develop as the temperature is lowered towards $T_{\rm KT}$ for
dopings where the superconducting phase is approached from the PG phase with
dopings which do not have a PG.

If the dynamics is described by the TDGL form Eq.~(\ref{eq:3}), then the peak
in $\Delta I(V,T)$ will occur at a voltage which varies as
$\varepsilon(T)=1-\lambda_d(T)\approx\chi^{-1}_d(T)$ at low temperatures.
Results for $\varepsilon(T)=1-\lambda_d(T)$ are shown in the Supplemental
Masterial. However, even if the dynamic structure of Im$\chi_d(\omega)$ is not
adequately described by Eq.~(\ref{eq:3}) \cite{ref:Koren,ref:Janko,ref:She},
the voltage integral of $\Delta I(V,T)/V$ will be proportional to $\chi_d(T)$.
Plots of $\chi^{-1}_d(T)$ are shown in Fig.~\ref{fig:4} for various dopings.
The inset in Fig.~\ref{fig:4} shows Monte-Carlo results for the inverse spin
susceptibility $\chi^{-1}(T)$ of the classical 2D xy model. Here one sees that
there is a Curie-Weiss regime at higher temperatures associated with
Emery-Kivelson phase fluctuations \cite{ref:EK} which then crosses over to the
low temperature vortex-antivortex KT behavior.
\begin{equation}
  \chi^{-1}(T)\sim\exp(-\frac{b}{\sqrt{T/T_{KT}-1}})
\label{eq:8}
\end{equation}
as $T_{\rm KT}$ is approached. 

We believe that the change in curvature of $\chi^{-1}_d(T)$ as the temperature
decreases for the $\langle n\rangle=0.95$ and 0.93 dopings reflects the onset
of phase fluctuations \cite{ref:EK} as $T$ decreases. This behavior is
analogous to that of a granular superconductor in which at higher temperatures
one has a BCS $\log(T/T^{\rm MF}_c)$ behavior associated with a single grain
followed by an xy Curie-Weiss behavior associated with pair phase fluctuations
for a range of temperatures until the KT behavior is reached. We note that
this change in curvature and the upturn at low temperatures is not seen in DCA
calculations using a 4-site cluster (2$\times$2 -- plaquette). In addition, in
this case, DCA calculations find that $T_c(x)$ has a maximum for $\langle
n\rangle=0.95$ and falls to zero very close to $\langle n\rangle=1$
\cite{ref:JarrellEPL01}, i.e. different from the 12-site cluster results
displayed in Fig.~\ref{fig:2}. We believe that this is due to the fact that
(spatial) phase fluctuations and KT behavior, which reduce $T_c$, are absent in
small clusters. This characteristic change in behavior as the cluster size is
increased provides further support for the presence of phase fluctuations in the
underdoped PG region of the Hubbard model. 

In contrast, for $\langle n\rangle=0.875$ and 0.85 the superconducting
transition is approached from a region without a PG. In this doping regime we
expect that the meanfield temperature $T^{\rm MF}_c$ is close to the
Kosterlitz-Thouless temperature and that over most of the temperature range
above a narrow region, set by the Ginzburg parameter, $\chi^{-1}_d(T)$ has the
GL form $\ln(T/T^{MF}_c)$. The data shown in Fig.~\ref{fig:4}(c) and (d) are
consistent with this behavior. 


Finally, although it is difficult to experimentally measure the large $q$
pairfield fluctuations \cite{ref:Koren} which are necessary to determine the
short distance pairfield susceptibility, in the numerical simulations this can
be done. With
$\Delta^\dagger_{\ell+x,\ell}=c^\dagger_{\ell+x\uparrow}c^\dagger_
{\ell\downarrow}$ creating a pair on site $\ell$ and its next near neighbor
site in the $x$ direction $\ell+x$, we have calculated the local $\chi_{yx}(T)$
pairfield susceptibility
\begin{equation}
  \chi_{yx}(T)=\frac{1}{N}\sum_\ell\int_0^\beta d\tau\langle\Delta^
  {\phantom\dagger}_{\ell+y,\ell}(\tau)
	\Delta^\dagger_{\ell+x,\ell}(0)\rangle
	\label{eq:11}
\end{equation}
This measures the local pairfield induced on the $(\ell,\ell+y)$ link when a
singlet pair is created on the adjacent $(\ell,\ell+x)$ link. It's negative
sign clearly shows the $d$-wave character of the local pairfield. We have
chosen to study $\chi_{yx}(T)$ rather than the local $d$-wave susceptibility
because $\chi_{yx}(T)$ avoids a remnant of the equal time expectation value
$\langle\Delta^{\phantom\dagger}_{\ell+x,x}\Delta^\dagger_{\ell+x,\ell}\rangle=
-2\langle{\bf
s}_{\ell+x}\cdot{\bf s}_\ell\rangle+\frac{1}{2}\langle
n_{\ell+x}n_\ell\rangle$ which is associated with the local spin and charge
correlations.

\begin{figure}[htbp]
\includegraphics[width=8.5cm]{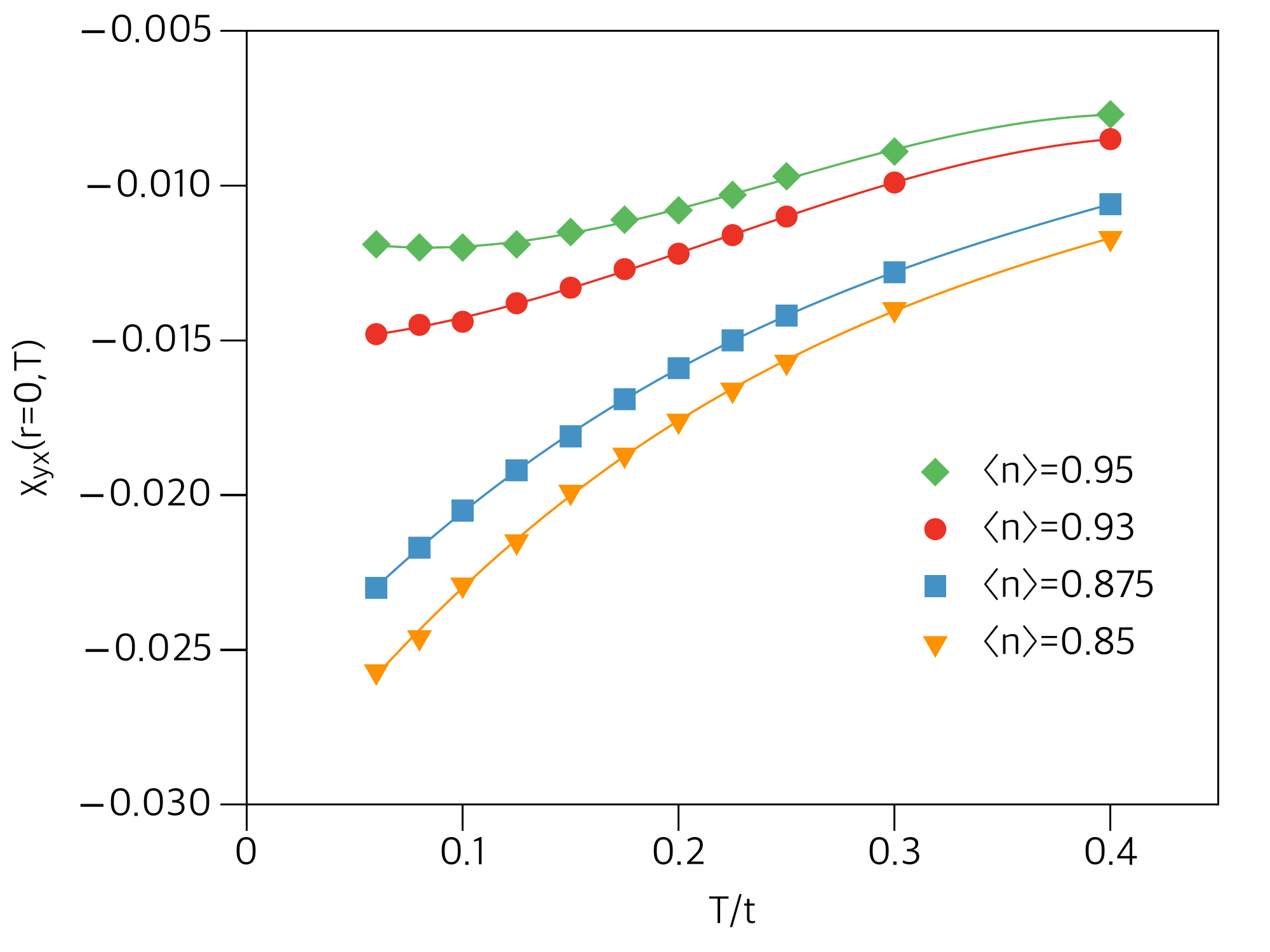}
  \caption{(Color online) The local $\chi_{yx}(T)$ pairfield susceptibility
  versus temperature $T$ for different fillings. The negative sign reflects the
  $d$-wave nature of the pairfield correlations. In the absence of a PG, these
  correlations continue to increase as the temperature decreases, while if
  there is a PG, they saturate.} \label{fig:5}
\end{figure}

Results for $\chi_{yx}(T)$ are shown in Fig.~\ref{fig:5}.
For the larger dopings, the local $\chi_{yx}(T)$ pairfield
susceptibility grows as $T$ decreases. However, for the underdoped cases,
$\chi_{yx}(T)$ saturates as the temperature decreases below $T^*$ and the
system enters the PG regime. In this case the amplitude of the induced local
pairfield is limited by the opening of the PG.

To conclude, the temperature dependence of the $d$-wave pairfield
susceptibility at the larger doping is consistent with Ginzburg-Landau Gaussian
amplitude fluctuations of the pairfield. At smaller dopings where the
superconducting state emerges at lower temperatures from a PG phase, there is
evidence that the growth of the local pairfield is limited by the opening of the
PG and the increase of $\chi_d(T)$ is associated with the development of long
range phase coherence.


\section*{Acknowledgments}

The authors would like to thank E.~Abrahams, I.~Bozovic, I.~Esterlis,
M.~Fisher, S.~Kivelson, P.A.~Lee, and A.-M.~Tremblay for useful comments. We
also thank I.~Esterlis for the xy Monte Carlo results shown in
Fig.~\ref{fig:4} and S.~Kivelson for his comment regarding the relationship of
this work to granular superconducting films.
This work was supported by the Scientific
Discovery through Advanced Computing (SciDAC) program funded by U.S.
Department of Energy, Office of Science, Advanced Scientific Computing
Research and Basic Energy Sciences, Division of Materials Sciences and
Engineering. An award of computer time was provided by the INCITE program.
This research used resources of the Oak Ridge Leadership Computing Facility,
which is a DOE Office of Science User Facility supported under Contract
DE-AC05-00OR22725.



\end{document}


\newcommand{\etal}{{\it et al.}\/}
\newcommand{\gtwid}{\mathrel{\raise.3ex\hbox{$>$\kern-.75em\lower1ex\hbox{$\sim$}}}}
\newcommand{\ltwid}{\mathrel{\raise.3ex\hbox{$<$\kern-.75em\lower1ex\hbox{$\sim$}}}}

\title{Supplemental Material: Pairfield Fluctuations of the 2D Hubbard Model}

\author{T.~A.~Maier}
\affiliation{Computational Sciences and Engineering Division, Oak Ridge
National Laboratory, Oak Ridge, Tennessee 37831-6164, USA}

\author{D.~J.~Scalapino}
\affiliation{Department of Physics, University of California, Santa Barbara,
CA 93106-9530, USA}

\date{\today}

\maketitle




\section*{Temperature dependence of the inverse pair life-time}

If the dynamics of the pairfield is described by Eq.~(3), then the voltage peak
$V_p(T)$ of the excess current $\Delta I(V,T)$ will be proportional to
$\varepsilon(T)=1-\lambda_d(T)$. In this relaxation approximation, one can think
of $\varepsilon(T)$ as proportional to the inverse life-time of a pair. In
Fig.~S1, we have plotted $\varepsilon(T)$ versus $T$ for the various dopings
discussed in this paper. Here on sees a similar behavior to the $\chi_d^{-1}(T)$
plots of Fig.~4. However, without the influence of the numerator in Eq.~(6), the
difference in behavior between the fillings which have a PG and those which do
not becomes more evident. 

\begin{figure}[htbp]
\includegraphics[width=12.0cm]{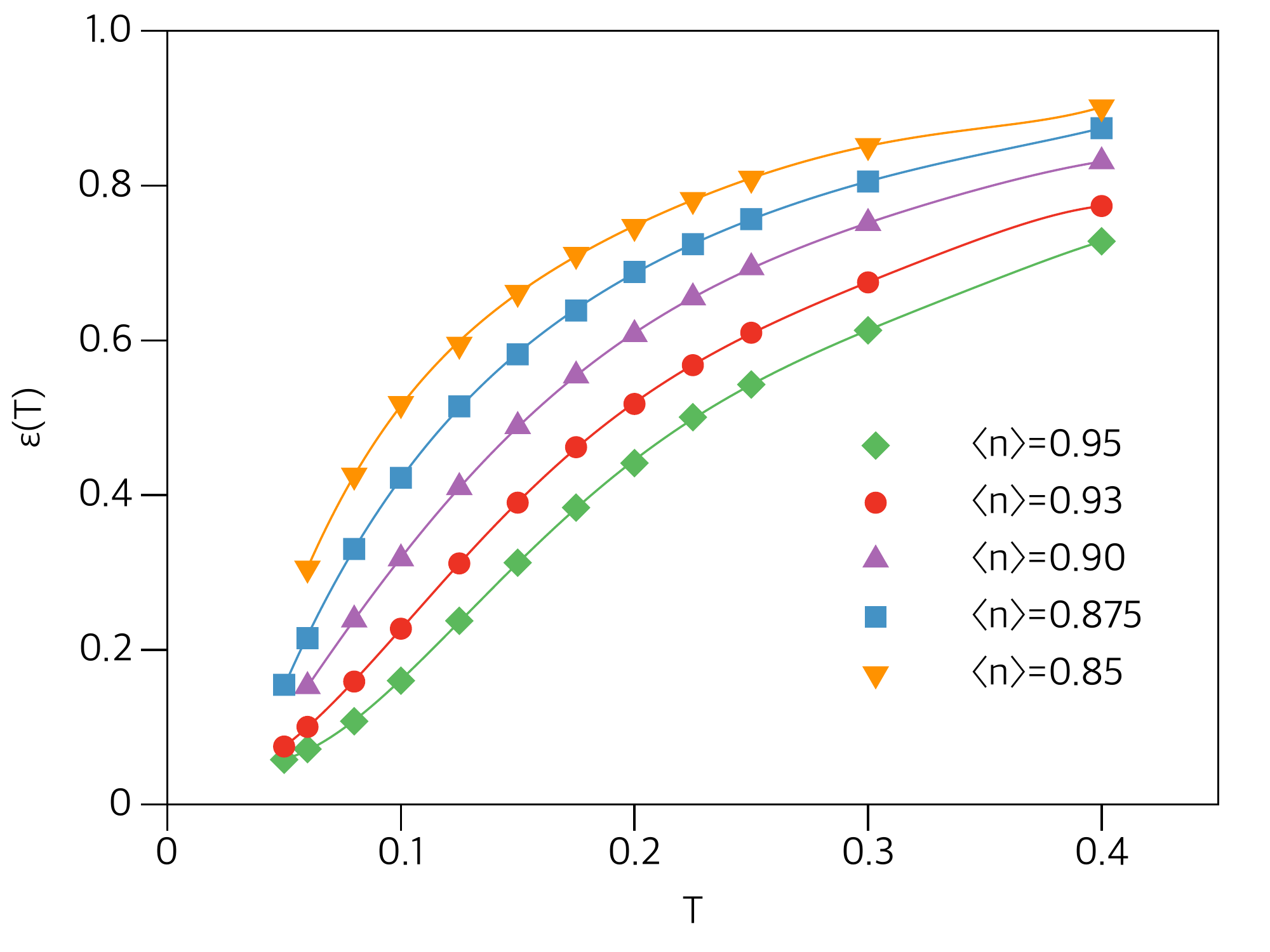}
  \caption{(Color online) Temperature dependence of $\varepsilon(T)=1-\lambda_d
  (T)$, where $\lambda_d(T)$ is the $d$-wave eigenvalue of the Bethe-Salpeter
  equation (8) in the 2D Hubbard model, for the various dopings discussed in
  this paper. In the underdoped region with a PG, the change in curvature
  arises from phase fluctuations, which merge with KT behavior at lower
  temperatures, while for larger dopings, the behavior is associated with
  Gaussian amplitude fluctuations. 
  \label{fig:S1}}
\end{figure}

In terms of the pair life-time, one can say that for
the fillings without a PG there is a range of temperatures above the KT
vortex-antivortex regime, in which this time is associated with a dissociation
of the quasiparticles which make up a pair. Alternatively, for fillings with a
PG, the lifetime is a measure of the phase coherence time. Here, for a range of
temperatures, these can be thought of as Emery-Kivelson phase fluctuations,
which, as $T_{\rm KT}$ is approaches for this 2D system, become associated with
vortex-antivortex fluctuations.